\documentclass[prl,onecolumn,noshowpacs,preprintnumbers,amsmath,amssymb]{revtex4}
\usepackage{graphicx}
\usepackage{color}
\usepackage{amsmath}
\usepackage{pifont}
\usepackage{epstopdf}
\usepackage[T1]{fontenc}
\usepackage{subfig}
\usepackage{float}

\begin{document}

\title{Universal Three-Dimensional Optical Logic}

\author{Logan G. Wright$^*$, William H.  Renninger, and Frank W.  Wise}
\affiliation{School of Applied and Engineering Physics, Cornell University, Ithaca, New York 14853}
\address{$^*$Corresponding author: lgw32@cornell.edu}

\begin{abstract}
Modern integrated circuits are essentially two-dimensional (2D). Partial three-dimensional (3D) integration\cite{Topol2006} and 3D-transistor-level integrated circuits\cite{Patti2006} have long been anticipated as routes to improve the performance, cost and size of electronic computing systems\cite{Akasaka1986}. Even as electronics approach fundamental limits however, stubborn challenges in 3D circuits\cite{Akasaka1986}, and innovations in planar technology have delayed the dimensional transition. Optical computing offers potential for new computing approaches, substantially greater performance and would complement technologies in optical interconnects\cite{Miller2000} and data storage\cite{Cumpston1999}. Nevertheless, despite some progress, few proposed optical transistors possess essential features required for integration into real computing systems\cite{Mcleod1995,Miller2010}.  Here we demonstrate a logic gate based on universal features of nonlinear wave propagation: spatiotemporal instability and collapse. It meets the scaling criteria and enables a 3D, reconfigurable, globally-hyperconnected architecture that may achieve an exponential speed up over conventional platforms. It provides an attractive building block for future optical computers, where its universality should facilitate flexible implementations.
\end{abstract}
\maketitle

2D and 3D electronic circuits are limited to roughly $2^{aN}$ configurations (a = interconnections/logic gate = 2-30, $N$ = number of gates. See Supplementary Discussion I for details.). Optics however, may enable unique and powerful new architectures\cite{Woods2012}. Spatially coherent light may transmit rapidly over long distances, without waveguides or wires, allowing for efficient wireless (i.e., continuously-connected), communication among all $N$ gates. This allows for roughly $2^N$  times more configurations. Such an architecture could be easily reconfigurable and ultimately more analogous to the brain than to a modern microchip. Like the brain, increased connectivity can enhance computational capacity. With large $N$ and apt programming, the effective speed boost could be staggering. 

	Mcleod et al. identified one route to such a computing hardware. The particle-like interactions between 3D optical spatiotemporal solitons (STS) could be the basis of logic gates\cite{Mcleod1995}. Use of a soliton-dragging configuration allows 3D gates that remain one of only a few concepts\cite{Mcleod1995,Miller2010,Demircan2011,Lentine1989} that fulfill all the transistor criteria\cite{Mcleod1995,Miller2010}, and the only to do so in a homogenous, continuous 3D geometry. Experimental generation of 3D STS is difficult, and has only been reported in glass patterned with an array of waveguides\cite{Minardi2010}. A 3D computer will require long-lived STS in a homogeneous material. This remains an important open problem. 

	Here, we take inspiration from STS logic but instead use wave collapse and spatiotemporal instability --- ubiquitous processes that counter formation of STS --- to create a robust, 3D universal optical logic gate. This gate meets all fundamental criteria, can be implemented in most optical media and is an important first step towards a powerful computing platform to succeed electronic circuits.

\begin{figure*}[htb] 
\centerline{
\includegraphics[width=16.0cm]{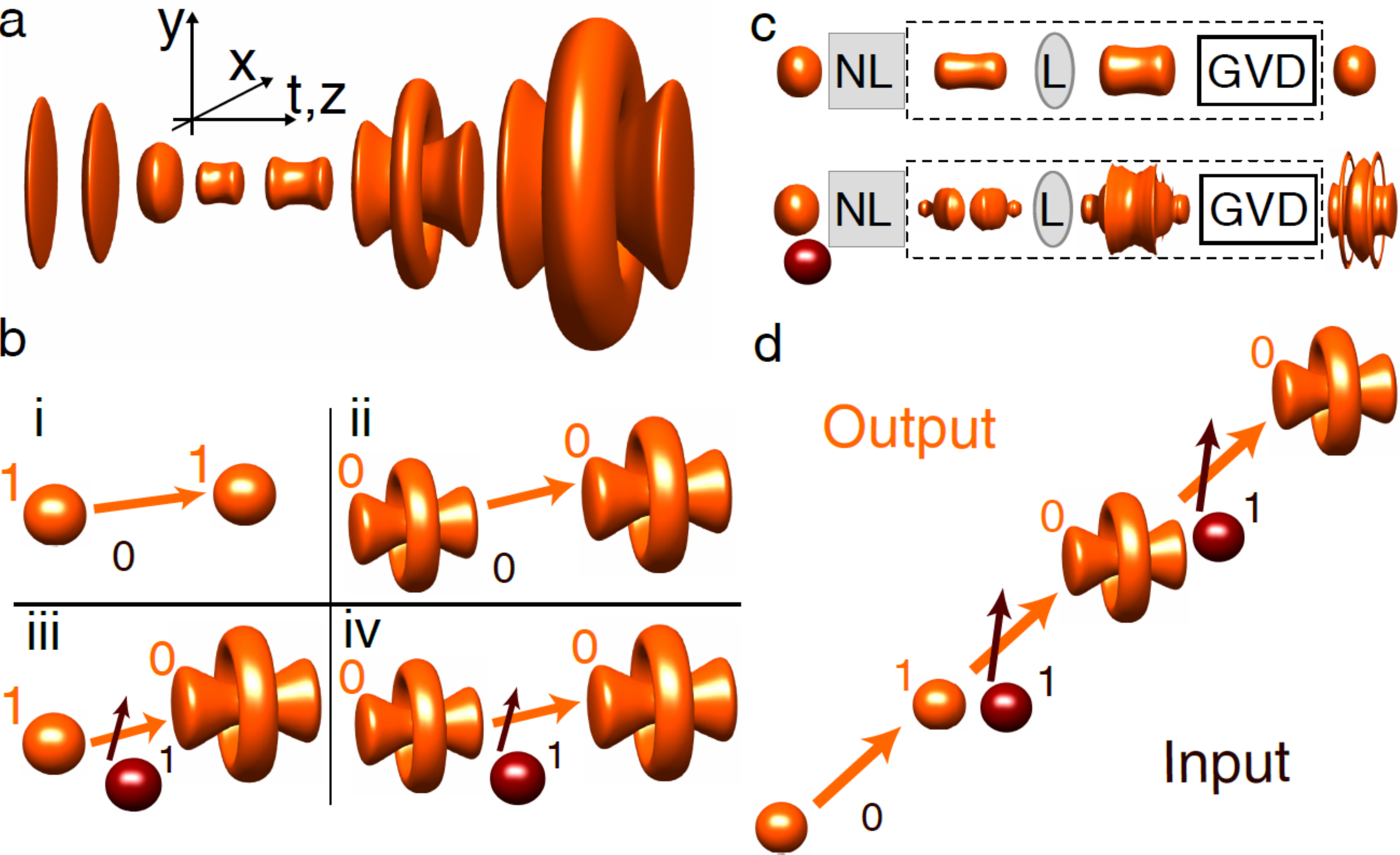}}
\caption{\textbf{Conceptual outline of optical wave collapse,  spatiotemporal instability, and use in 3D optical computing.} \textbf{a,} Isosurfaces at 10\% maximum intensity for an ultrashort pulse undergoing self-focusing followed by spatiotemporal pulse splitting. \textbf{b,} Transistor operations: a Gaussian (round) wave corresponds to a 1, a post-collapse (conical) wave to a 0. i. without signal, 1 remains 1; ii. without signal, 0 remains 0; iii. with signal, 1 transforms to 0; iv. with signal, 0 remains 0. \textbf{c,} Depiction of linear regeneration (NL: nonlinear medium, L: linear optics (linear propagation, lens), GVD: dispersion compensation).  Restoration of the pulse's spatiotemporal intensity profile is only possible if the wave has not undergone collapse (top). \textbf{d,} Schematic of a 3-input NOR gate with a (011) input. }
\label{sym2d}
\end{figure*}

An intense optical beam in a transparent material collapses towards a point due to the optical Kerr effect. Any initial optical field distribution (provided only that it possesses slightly more than a material-dependent threshold power, $P_{crit}$), will self-focus and evolve asymptotically towards the circularly-symmetric Townes profile\cite{Moll2003}. The Townes profile is unstable\cite{Moll2003,Porras2007} and collapse is inevitably arrested by other processes. In transparent media the dispersion is typically normal, and nonlinearity and dispersion may act together to split a pulse in space and time (Fig. 1a,  ref. 14). This irreversibly transforms it into a post-collapse wave, whose signature feature is the presence of conical structures, in which angle and frequency are proportional, surrounding a spectrally-broad (i.e., supercontinuum) circular core. The details of collapse are the subject of extensive research\cite{Moll2003,Porras2007,Rothenberg1992,Faccio2006,Conti2004,BoydSF,Gaeta2000,Ishaaya2007}. An important feature of the logic operation that we describe below is that it is independent of these details, relying only on the universal (occurring in nearly all materials) transformation through collapse.  

Collapse and its subsequent arrest is an irreversible process: a metastable Gaussian wave with P>$P_{crit}$ is transformed into a stable, extended post-collapse wave (Fig. 1a). The gate operation may be viewed in analogy with a spontaneous combustion reaction. While nominally stable, a combustible material undergoes a marked transformation if ignited or exposed to a certain critical reagent or catalyst. This ignition acts to lower the potential barrier separating the metastable initial state from the lowest energy state, causing a spontaneous reaction. In the same way, we here demonstrate that a weak pulse (the signal pulse) may induce the spontaneous transformation of a stronger pulse. This is the central advance of the present work: the demonstration of a controlled transformation which can be used as an amplifying switch (a transistor) that satisfies all the essential requirements for scalable computing. Via the Kerr effect, distinct fields may exert optical forces on one another. As suggested by theoretical analysis\cite{Berge1998} and shown experimentally here, these coupling forces allow even a weak secondary beam to accelerate the collapse of a stronger one, ''igniting'' the runaway collapse process. At the point of collapse, high intensities allow excitation of additional spatiotemporal instabilities, through which the pump field is transformed into the most stable state: the post-collapse wave.

 The possible interactions of a weak signal pulse with a strong pump pulse are illustrated in Fig. 1b. While the signal may trigger the transformation of the Gaussian-like pump (iii), it cannot induce transformation of a post-collapse wave (iv). If the pump does not collapse in the gate, linear regeneration (via linear optics, linear propagation, and dispersion compensation) can adequately restore the pump to its initial condition. For a post-collapse wave, these measures will have no restorative effect, so the difference between the two states will be amplified (Fig. 1c). By directing the output pump pulse into a subsequent nonlinear medium, a NOR gate may be constructed.  Repeating the process n times produces an n-input NOR, which is a functionally complete operation (Fig. 1d). As the pump propagates, it interacts with signal pulses whose presence corresponds to an input logical 1. The pump may be efficiently switched by a single input beam and any additional beams will not undo this. As a result, the NOR gate yields the correct logical 0 unless all inputs are 0 (i.e., absent).

We chose to design the experiments to elucidate the universal characteristics of the interactions and so conducted experiments in long (3x10 cm) glass rods, using 300-fs pulses of light with up to 1 $\mu$J total energy. These experiments are not intended to directly show scalability, but to provide a simple, familiar setting to evaluate the new concept. The particular system allows us to evaluate universality, as the dynamics are well-described by the lowest-order nonlinear wave effects contained within the 3D nonlinear Schr{\"o}dinger equation (3D-NLSE). We expect qualitatively-similar results in most other materials (including those where the requisite length and energy will be vastly reduced), a conclusion which would be less definitive in other systems. 
Using a non-collinear, cross-polarized interferometer (Fig. S1) to allow control of the pulses' relative energy, angle, and time delay, we focus two pulses onto the front face of a rod of SF11 glass. When the two pulses have equal power and are overlapped temporally, a significant deflection is observed as the pulses interact similarly to solitons (Fig. 2a-b). However, gain is a prerequisite for a switch to be useful within a computer, since each gate must provide fan-out and compensate loss. As the energy ratio (gain) between the pump and signal is increased, these near-soliton interactions become weaker. We find, however, that the presence of a less-energetic pulse can trigger the collapse and spatiotemporal splitting of a stronger pump pulse. For example, with a gain of 5, the signal transforms the pump to a characteristic post-collapse profile (Fig. 2c-d). 

To understand the interactions, the 3D-NLSE was numerically solved assuming transform-limited input pulses with parameters close to the experimental values. The pump beam profiles (Fig. 2e-f) agree well with the corresponding experimental results (Fig. 2c-d).  Several additional comparisons, sampling the accessible parameter space, are shown in Fig. S2. The good agreement between experimental results and simulations that include only the lowest-order effects is important, as it strongly implies universality.

\begin{figure}[htb]
\centerline{
\includegraphics[width=8.0cm]{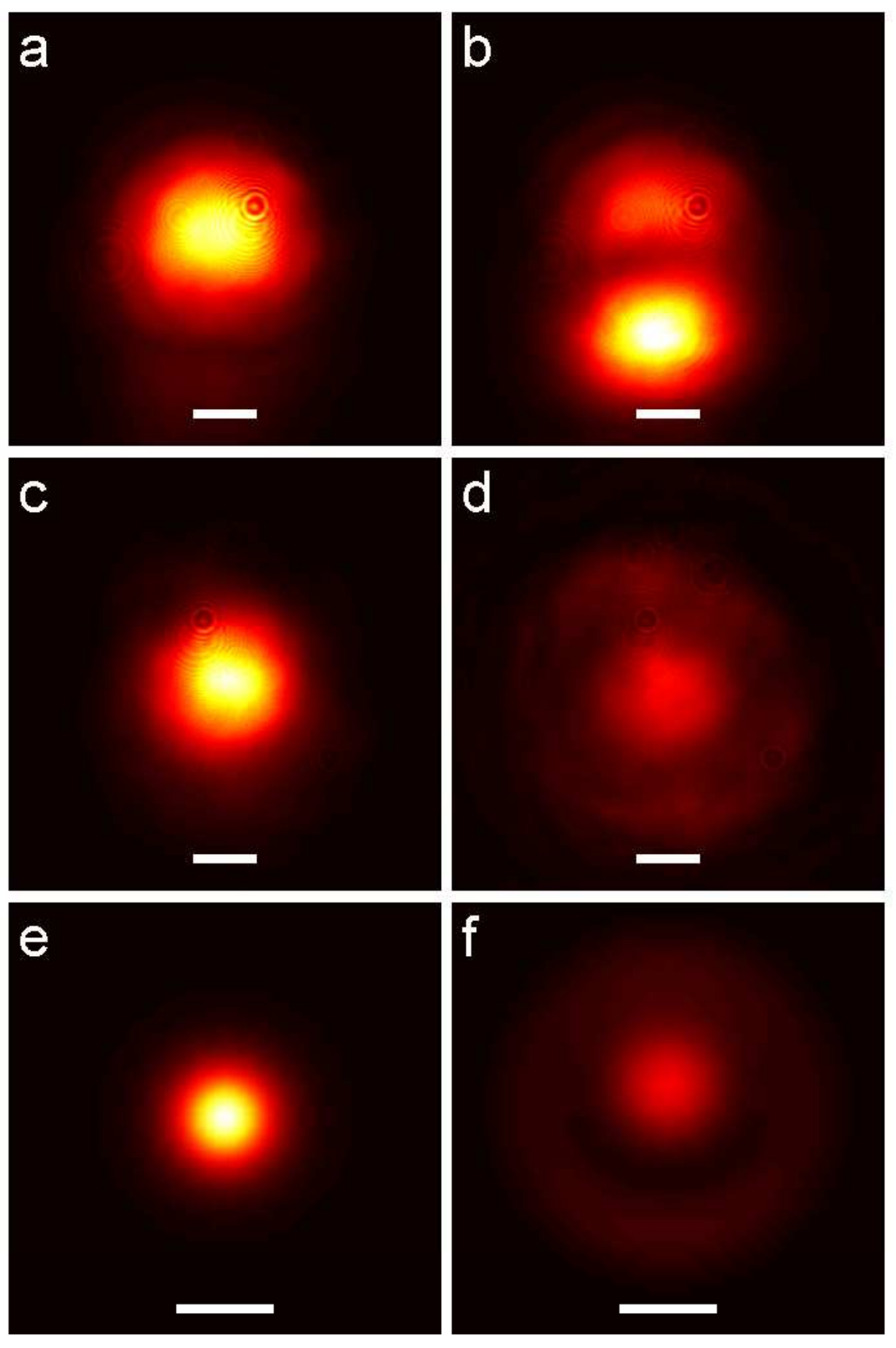}}
\caption{\textbf{Experimental observations of signal-controlled interactions in SF11 and simulations using the 3D-NLSE.} \textbf{a,} Beam profile of pump without, and \textbf{b} with signal (gain = $E_{pump}/E_{signal} = 1$). \textbf{c,} Pump beam profile without, and \textbf{d} with signal present (gain = 5). \textbf{e,} Simulated pump profile (1.25 MW peak power) without, and \textbf{f} with signal present. a and b are plotted with self-normalized color scales while c and d (e and f) are both normalized to the maximum in c (e). Scale bars correspond to 200 $\mu$m. 
}
\label{symbr}
\end{figure}

To demonstrate 3D scalability, we consider a 3D, 3-input NOR gate. The pump pulse is centered between three signals, which are arranged on the vertices of an equilateral triangle (Fig. 3, left). This realizes Fig. 1d but uses the freedom of three dimensions to cascade the gate within a smaller volume and with zero latency.  Numerical simulations of this arrangement yield results similar to the one (Fig. 2c-f) above, with the pump for each input case shown in Fig. 3, right. Further scaling is possible by placing additional signal beams within a circle centered on the pump. As shown in the Fig S3-4 and Supplementary Discussion II, we find that, by adjusting the signal geometry, this gate may simultaneously provide other logic functions, including 3D-NAND, and facilitate signal-controlled pump steering. Combining conventional binary-state logic with conditional steering and continuous, many-input gates should simplify system design and may provide more sophisticated computational functions.

\begin{figure*}[htb] 
\centerline{
\includegraphics[width=4.3cm]{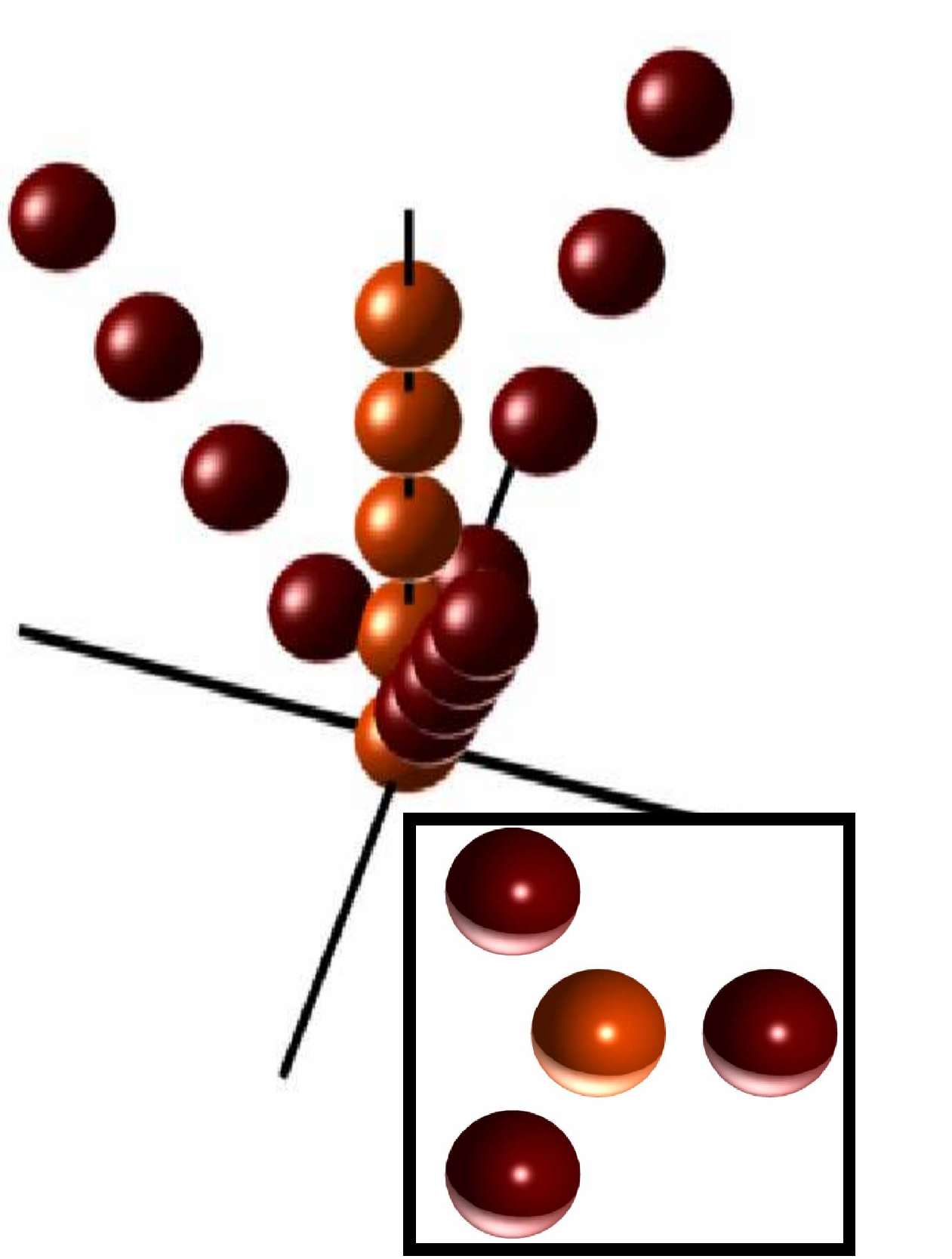}
\includegraphics[width=11.7cm]{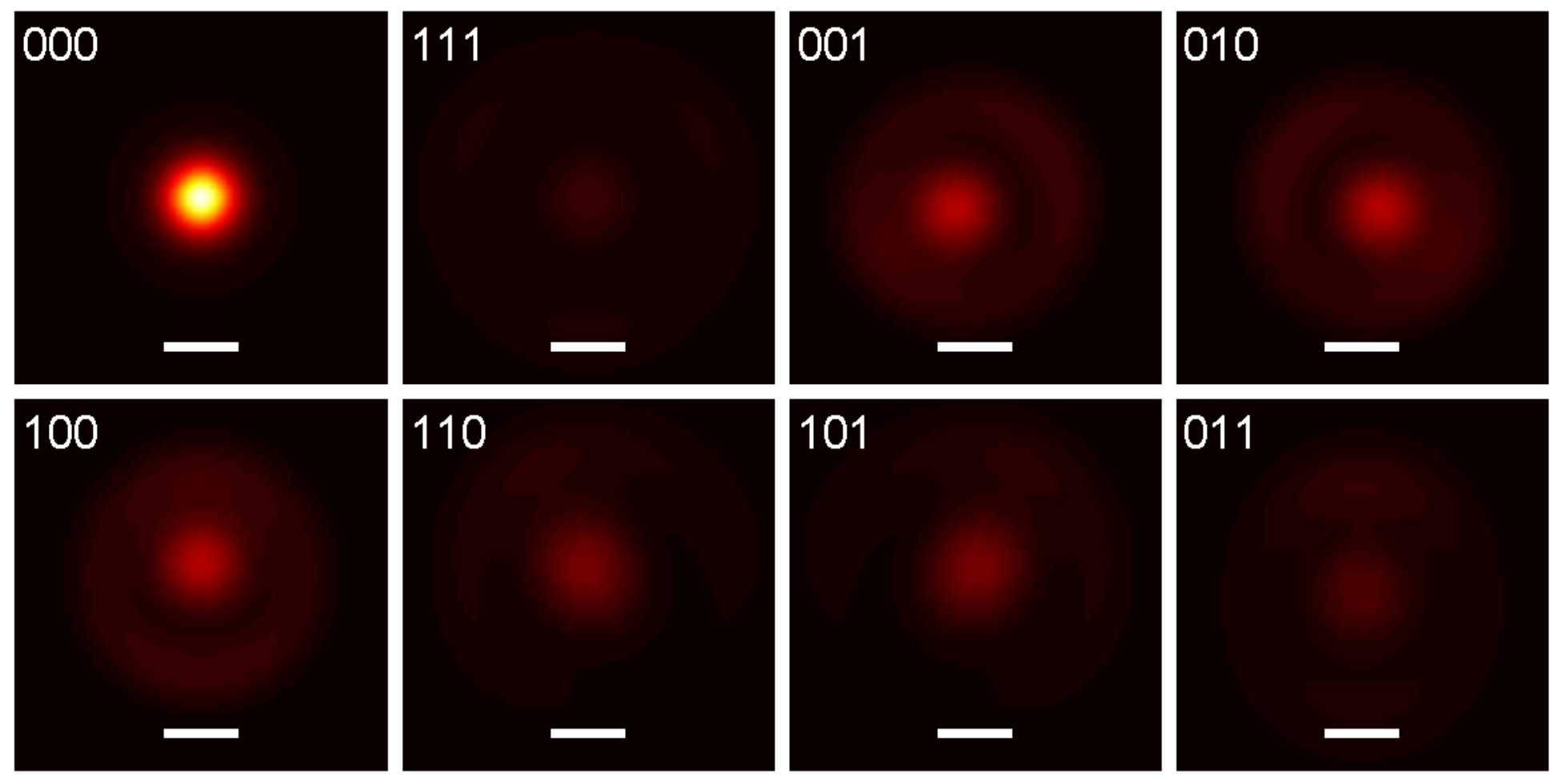}}
\caption{\textbf{Three-dimensional, 3-input NOR gate simulation.} \textbf{left} illustration of the interaction geometry (pump orange, signal red). Inset: cross-sectional view. \textbf{right} pump profile for each possible logical input. Numbers indicate the presence of each of the three signals, symmetrically arranged (120 degrees between each) around the pump (1.3 MW) at an angle of 1.2 mrad. Gain is 5. Scale bars correspond to 200 $\mu$m. }
\end{figure*}

If properly designed (see Supplementary Discussion III), logic gates based on the induced instability of the pump pulse possess all necessary qualities required for integration into a full-scale computer\cite{Mcleod1995,Miller2010}. The output pulse of one gate can drive a subsequent gate, and because they provide gain, there is capacity for fan-out. The insensitivity of evolution towards the Townes profile to initial conditions intrinsically damps fluctuations. The gates provide logic level restoration (fluctuations in input parameters do not result in fluctuations of output level) and do not require critical biasing. Indeed, although the laser in the experiments described here exhibits up to 20\% energy fluctuation, the gate operations of Fig. 2 are stable and highly reproducible. The irreversibility of the collapse process enforces unidirectional errors, which will allow efficient error correction.  The use of orthogonal circularly-polarized fields in an inverting logic operation eliminates phase-dependence and provides input-output isolation. Electrical detection schemes that are independent of path-specific loss may be conceived; most multi-point measurements of phase, spectrum or beam shape should work. In addition, the distinct characteristics of pre- and post-collapse waves imply that the logic contrast will be enhanced by extended propagation\cite{Shim2012,DiTrapani2003} (e.g., through subsequent gates). These criteria will be similarly satisfied with other physical realizations, where the transformation, or arrest of collapse, is mediated by effects other than Kerr nonlinearity and dispersion\cite{Shim2012,Dubietis2004}.
The exponential configuration boost enabled by the gates described here motivates development, but universality and optical advantages will be key in empowering it. Together, these features suggest caution in making direct comparison to the requirements of any previously-described systems. As an obvious step, the gates can be implemented in highly nonlinear materials, thereby lowering power and size requirements. Similar to the advantages the gates provide for software however, they afford many advantages for novel hardware designs that, while radically different from electronic designs, may facilitate practical devices.  Devices may consist largely of homogenous, linear dielectrics (even air), enabling low-loss, crosstalk-free continuous interconnection and control over the location of energy dissipation.  Quantum impedance conversion implies that only a minute sample of each pulse is required for electronic detection. Hence, energy-efficient reuse of '1' (and, if an efficient conversion scheme can be devised, '0') logic pulses should be possible over many computing cycles. These measures may combine for a device with very small net power dissipation in spite of the relatively high peak power required for gate operations. These illustrative example strategies merely hint at the architecture design possibilities. Specific measures may improve gates themselves. For example: the pump's propagation may be controlled\cite{Fibich2006,Cao1994,Li2012} to avoid accidental collapse, and the signal characteristics may be adjusted to more strongly excite instability\cite{Porras2007,Conti2004,Liou1992}. E.g., using a weak pulse to trigger supercontinuuum generation\cite{Ensley2011,Solli2008} has yielded gain > 100 and 30 dB contrast. Since the gates provide logical inversion, more immediate applications in signal processing should be possible\cite{Gao2008}.

We have demonstrated an approach towards a continuous, 3D, globally-interconnected, wireless optical computing architecture, based on the instability of optical waves in nonlinear media.  Complemented by emerging technologies in optical data storage and broadband interconnects, this opens a route to powerful computing technology. Its universality and robustness should translate to flexibility in design and implementation in practical systems. 

\section{Acknowledgments } 
Portions of this work were supported by the National Science Foundation (PHY-0653482). L.G.W acknowledges support of an NSERC PGS-D scholarship.  The authors thank A. Gaeta and K. Wagner for discussions.

\pagebreak
\renewcommand{\figurename}{\textbf{Supplementary Figure}}
\setcounter{figure}{0}

\title{Supplementary Information for 'Universal Three-Dimensional Optical Logic'}

\author{Logan G. Wright$^*$, William H.  Renninger, and Frank W.  Wise}
\affiliation{School of Applied and Engineering Physics, Cornell University, Ithaca, New York 14853}
\address{$^*$Corresponding author: lgw32@cornell.edu}
\maketitle

\begin{figure}[H] 
\centerline{
\includegraphics[width=16.0cm]{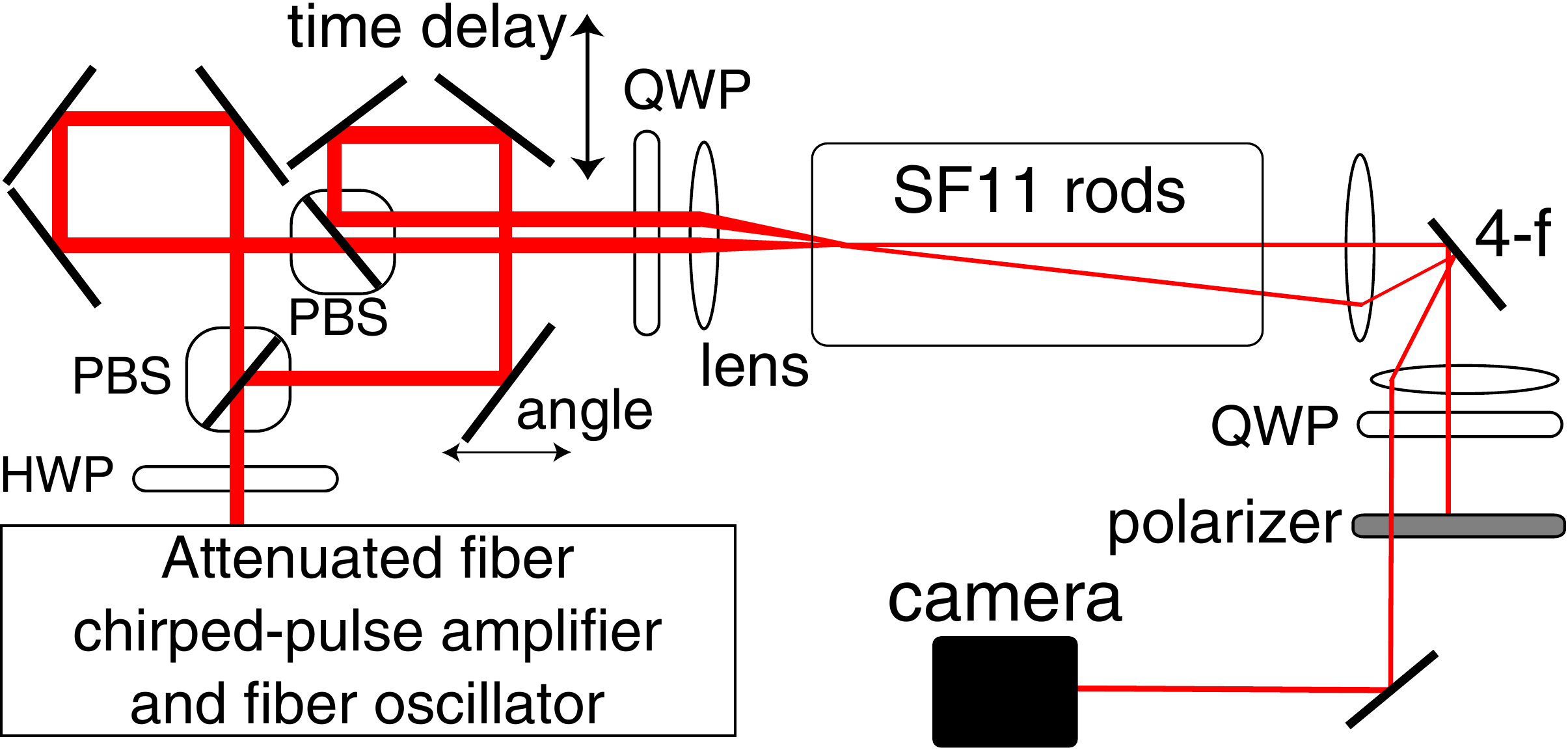}}
\caption{\textbf{Schematic diagram of experimental apparatus.} A source of 1 $\mu$J pulses is routed into a polarizing beam splitter (PBS), where the fraction travelling in either direction is controlled by the half-wave plate (HWP). The angle between the two copies of the pulse is controlled by adjusting the transverse offset of the beams prior to the final focusing lens (using the 'angle' translation stage). The time delay between the pulses is adjusted by translating the second and third mirrors in the reflected-polarization's path ('time delay'). The two copies are recombined on the second PBS and focused by a lens onto the front face of the SF11 rods. A quarter-wave plate (QWP) converts the beam polarizations to orthogonal circular polarization states. After the rods, the beams are imaged by a 4-f imaging system onto a beam profiling camera. The signal is filtered from the pump by a quarter waveplate and polarizer.}
\label{sym2d}
\end{figure}

\section{Methods}
\textbf{Experiment}
Pulses from a homemade, pulse-picked fiber oscillator seed a fiber-based chirped-pulse amplifier that produces roughly 320 fs full-width-at-half-maximum (FWHM) pulses with up to 1$\mu$J at 1030 nm. These pulses are then attenuated using a series of waveplates and polarizers and split in a setup resembling a polarizing, noncollinear Mach-Zehnder interferometer. Independent control of the reflected-polarization beam's position and time delay is achieved using the two orthogonal translation stages on one path of the interferometer. The pulses are focused to about 200 $\mu$m (beam waist, $1/e^2$ intensity) onto the front face of an ordinary SF11 glass rod. After propagation through the rods (3 10 cm rods with polished and anti-reflection coated end-faces), the beams at the final end-face are imaged onto a camera using a 4-f imaging system. SF11 exhibits normal dispersion at 1030 nm, prohibiting temporal solitons and lacks the saturating nonlinear effect necessary to stabilize spatial solitons. No change in loss is observed with increasing power. Quarter-wave plates before and after the glass are used to convert the pulses to orthogonal circular polarizations and then back to linear polarizations so that the pump beam can be imaged separately from the signal by use of a polarizer. By varying the waveplates and translation stages, we access a parameter space of total power, relative angle, and relative power. No interactions are observed for relative delays of more than about 300 fs. In order to control for possible interactions with a weak background and slow or nonlocal effects we chose to measure the interactions by varying the delay rather than simply blocking either of the beams, as was done in the simulations. A diagram of the apparatus is shown in Supplementary Figure 1.

\textbf{Simulations}
We were guided initially by two-dimensional simulations, which were observed to match experiment qualitatively along the plane containing the angle between the beams. The simulations involved a split-step Fourier solution of the three-dimensional nonlinear Schr{\"o}dinger equation for each polarization. The coupled nonlinear step used a 4th-order Runge-Kutta method. The fields were chosen to initially have flat radial and spectral phases, were 320 fs (FWHM) by 200 $\mu$m ($1/e^2$ intensity) and were initially overlapped. Parameters for SF11 were n=1.76, $\beta_2=1260$ fs$^2$/cm, and n$_2$=41$\times10^{-8}$ $\mu$m$^2$/W. For 3D simulations, the grid was 128 time points by 256 space points with a 750 $\mu$m step in the propagation direction. Each simulation, involving both fields traversing the rods, requires approximately 20 minutes using 4 cores of an Intel i7 computer. For simulations shown in Figure 1a, 1b and 1d, the pulse duration was shortened to 150 fs. This leads to a simpler, more identifiably cone-shaped wave which typifies the general form of post-collapse waves and is more suitable for an explanatory figure of the general concept. For Figure 1c, parameters mentioned above were used, in order to demonstrate the effect of linear regeneration on the pulses in the experiment and simulated gates.

\section{Supplementary Discussion I: Comparison of Computing Architectures}
To provide an approximate measure to allow us to compare different computational platforms, we will consider the number of logical states of a set of $N$ logic gates, which are each comprised of several transistors. Of course, we cannot fully calculate the performance of each platform without making more detailed assumptions about the programming, the rate of error, power consumption, clock speed, etc. In addition to different levels of connectivity and flexibility, different platforms offer different advantages at the software and hardware levels. Novel platforms provide an opportunity for new advances in these areas while necessarily eschewing many well-developed technologies and paradigms. 

In conventional platforms, computation and interconnection takes place on a two-dimensional plane and gates are connected to their nearest neighbours by wires of finite size and fixed location. We will assume, for simplicity, that there are 8 input wires to each gate. These 8 inputs allow for $2^8$ independent output states. For $N$ gates, there are therefore $8N$ inputs each clock cycle, for $2^{8N}$ states. For operations that involve $n$ sequential gate operations over $n$ cycles, there are $2^{8Nn}$ states for the two-dimensional set. This factor of $2^n$ applies, within our assumptions, equally to all architectures. For a 3D wired architecture, there are 18 additional neighbours to each gate (again, assuming $45^{\circ}$ and $0^{\circ}$ wires), for a total of 26 inputs. Similarly to the 2D wired architecture, there are $2^{26Nn}=9.5\cdot2^{8Nn}$ states for $N$ gates in $n$ cycles. 

For the globally-interconnected, wireless 3D architecture, each gate is directly connected to (up to) all $N$ other gates. All space between gates can be used, simultaneously by all signals, to connect to other gates. The number of states in $n$ cycles is $2^{N^2n}$. For a cutting-edge modern microprocessor transistor count of ~$5\cdot10^9$, the number of states for the assumed 2D architecture is about $2^n\cdot2^{8\cdot10^9}$. For the 3D, globally-interconnected wireless architecture, the requisite $N$ to match this is about $10^5$. Hence, a set of about $10^5$ gates can be compared to $10^9$ in the conventional architecture. More importantly for future applications, a 'chip' with $10^9$ gates may be comparable to a set of $10^{17}$ in a conventional one. This factor of improvement corresponds to a boost in computational capacity similar to that which has been obtained since Gordon E. Moore first predicted the exponential growth of integrated circuit density in 1965. While this measure is incomplete, it suggests some caution is due when making comparisons of technical performance requirements with conventional architectures (\textit{e.g.}, energy consumption or device size), all of which operate within a similar 2D wired configuration. 

\pagebreak

\begin{figure}[H] 
\centerline{
\includegraphics[width=16.0cm]{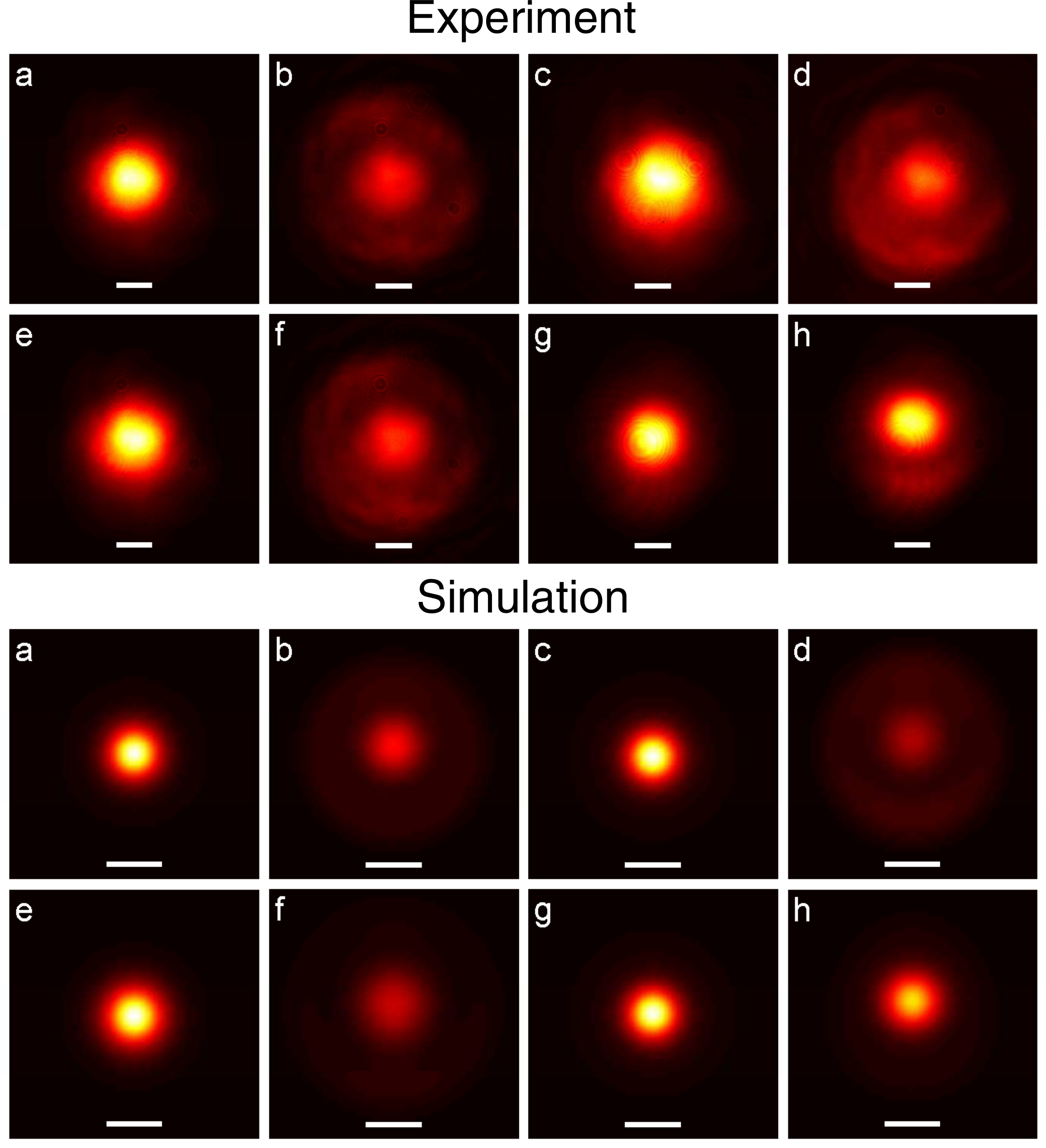}}
\caption{\textbf{Additional comparisons of experiment and simulation.}\textbf{Top}: arbitrarily selected experimental interactions. Pulse duration 320 fs, 200 $\mu$m beam waist. \textbf{a} Pump (1.32 MW) without signal,  \textbf{b} Pump (1.32 MW) with 0.15 MW signal at 0.7 mrad, \textbf{c} Pump (1.3 MW) without signal,  \textbf{d} Pump (1.3 MW) with 0.3 MW signal at 1.5 mrad, \textbf{e} Pump (1.2 MW) without signal,  \textbf{f} Pump (1.2 MW) with 0.3 MW signal at 0.9 mrad, \textbf{g} Pump (1.4 MW) without signal, \textbf{h} Pump (1.4 MW) with 0.28 MW signal at 2.4 mrad. \textbf{Bottom}: arbitrarily selected experimental interactions, simulated (parameters same as above). Excellent agreement is observed except for a systematic tendency for simulations to produce smaller beams. Simulations varying initial pulse chirp, radial phase and astigmatism suggest this is due to uncompensated nonlinear chirp in the amplified experimental pulses and because the slightly elliptical experimental beams possess an finite (and asymmetric) radial phase at the front face of the SF11. These factors were not found to influence the interactions in any significant way except to delay the onset of collapse. Scale bars correspond to 200 $\mu$m. }
\label{sym2d}
\end{figure}

\pagebreak

\section{Supplementary Discussion II: Other Logic Functions}

With a similar geometry as shown in Figure 3 of the letter, additional logic functions are possible. We assume similar parameters as in Figure 3 except as noted below. By increasing the angle of the signals, collapse of the pump may only be triggered with n>1 signals present. However, weak soliton-like dragging occurs unless symmetric signals are applied. As shown in Supplementary Figure 3 (and observed experimentally in Supplementary Figure 2g-h), these angular deflections may allow all-optical control of the pump without causing collapse. In a properly designed network, the gate in Supplementary Figure 3 may thus have a useful operation: depending on signal input, it either steers the pump, leaves it unchanged, or causes it to collapse. For strict NAND operation, we split signals symmetrically into to two copies, each containing one half the original signal energy. These copies cancel their respective induced deflections, leaving the pump beam's direction unaffected. For large angles, strict NAND operation is observed (Supplementary Figure 4). 

\begin{figure}[htb]
\centerline{
\includegraphics[width=13.0cm]{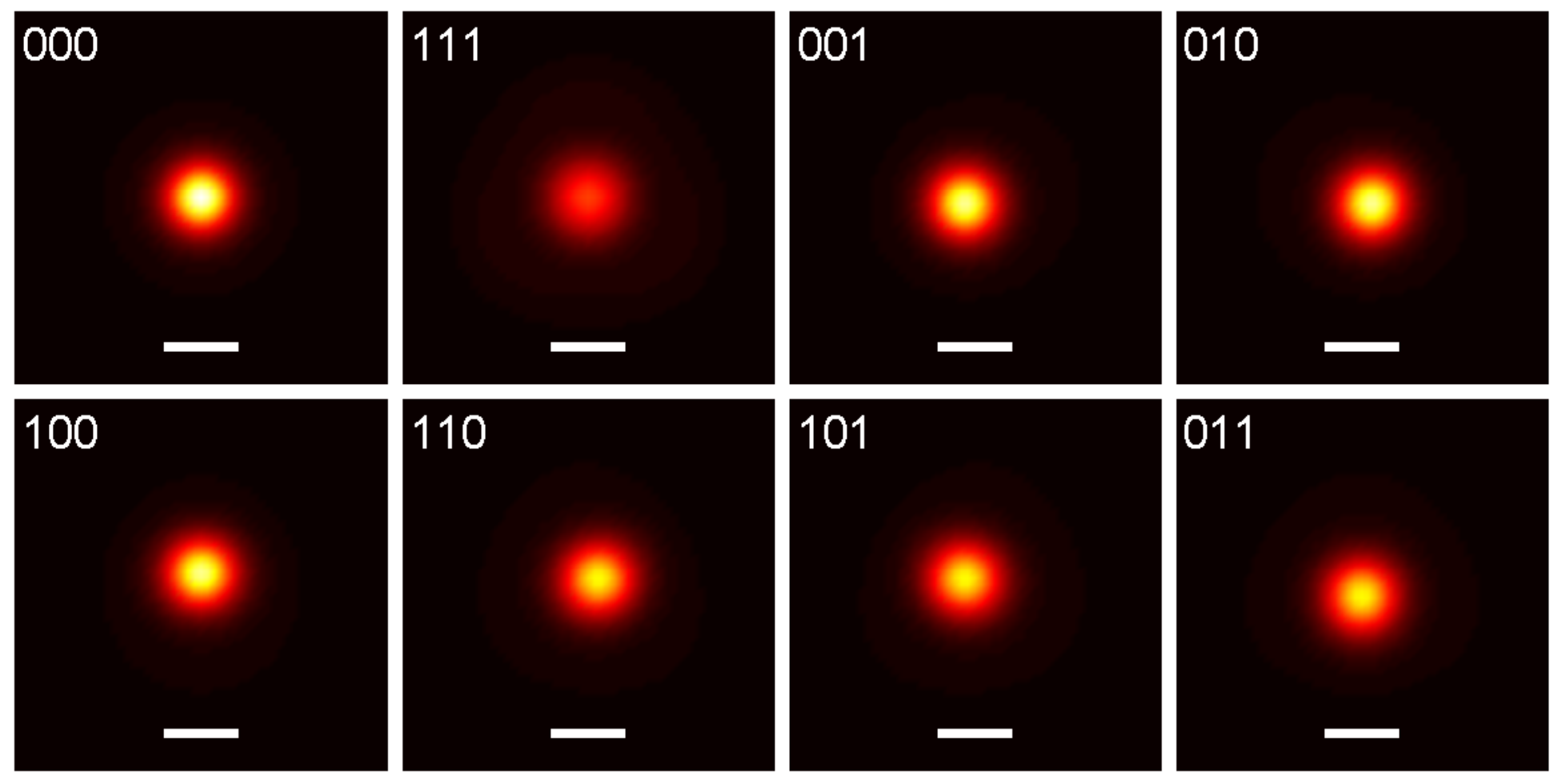}}
\caption{\textbf{Final spatial profile of pump beam for unbalanced 3D NAND.} Geometry of signals and pulse parameters are identical to Figure 3 in the letter except angle is 3 mrad. Small deflections occur, as in Supplementary Figure g-h. Scale bars correspond to 200 $\mu$m. 
}
\label{symbr}
\end{figure}

\begin{figure}[htb]
\centerline{
\includegraphics[width=13.0cm]{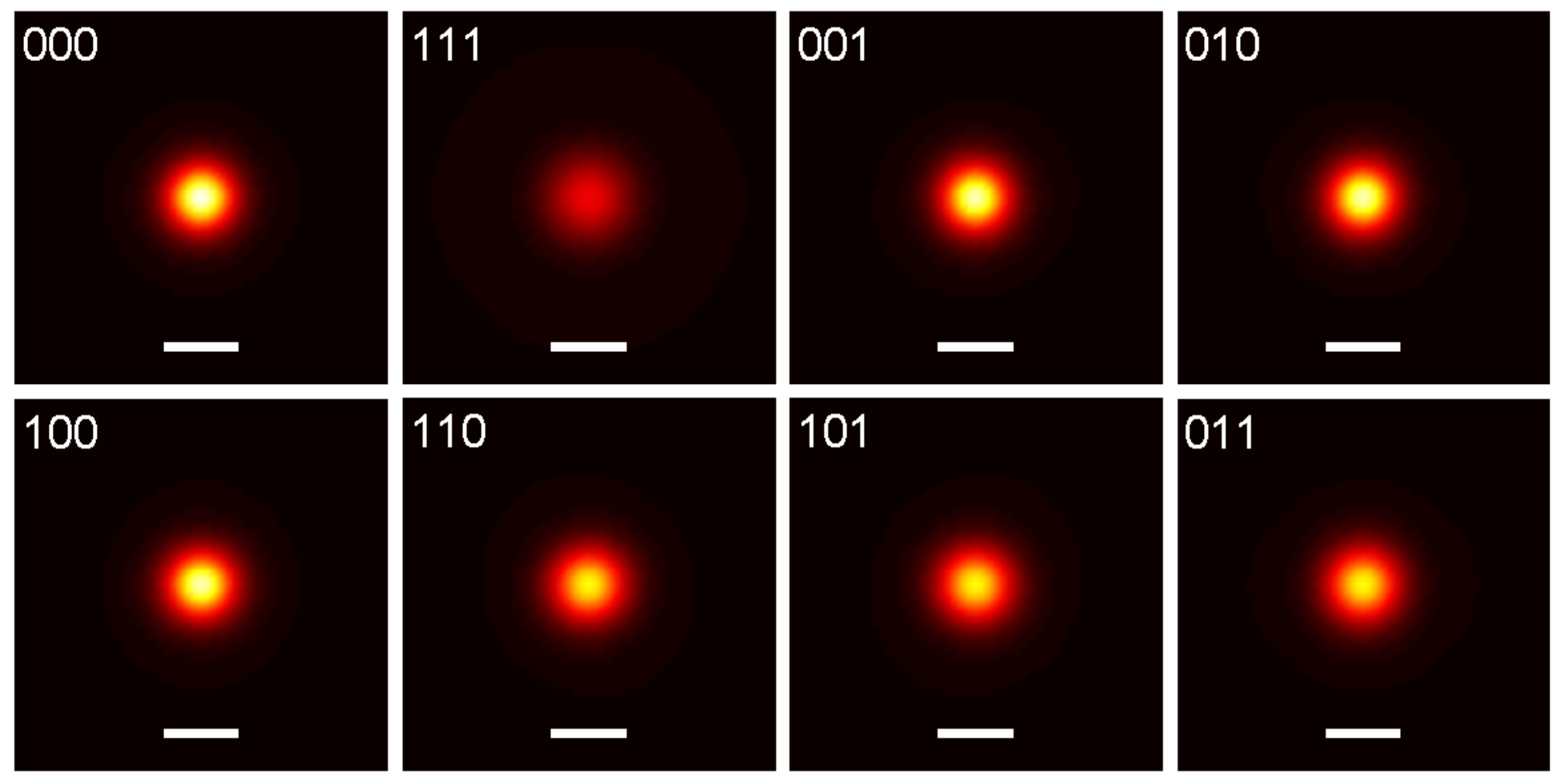}}
\caption{\textbf{Final spatial profile of pump beam for momentum-balanced 3D NAND.} Geometry of signals and pulse parameters are identical to Figure 3 in the letter except angle is 4 mrad and each signal is split so that two copies travel at opposite angles relative to the pump. This splitting avoids the deflections which occur in Supplementary Figure 3. Scale bars correspond to 200 $\mu$m. 
}
\label{symbr}
\end{figure}

\pagebreak

\section{Supplementary Discussion III: Gate Design and Operation}

A useful gate must maximize the effect of the signal's perturbation to the pump while simultaneously ensuring that the pump does not destabilize in solitary propagation and can be restored for routing into subsequent gates. To this end, we established an operating regime by increasing the pump power until the pump beam size was observed to increase in size with pump power. We then chose to restrict the pump power below the lowest power for which the minimum beam size could be observed. This value is expected to depend on the details of the collapse-arresting mechanisms, as well as on the initial pulse parameters (spot size, radial phase, pulse duration, chirp, etc.). Manipulation of these parameters is thus expected to allow tuning of device performance. Provided the pump power is set sufficiently below this point - the point where collapse occurs in the medium - gate operation should not be sensitive to noise. We have observed this insensitivity both experimentally and in simulation. 

\begin{figure}[htb]
\centerline{
\includegraphics[width=23.0cm]{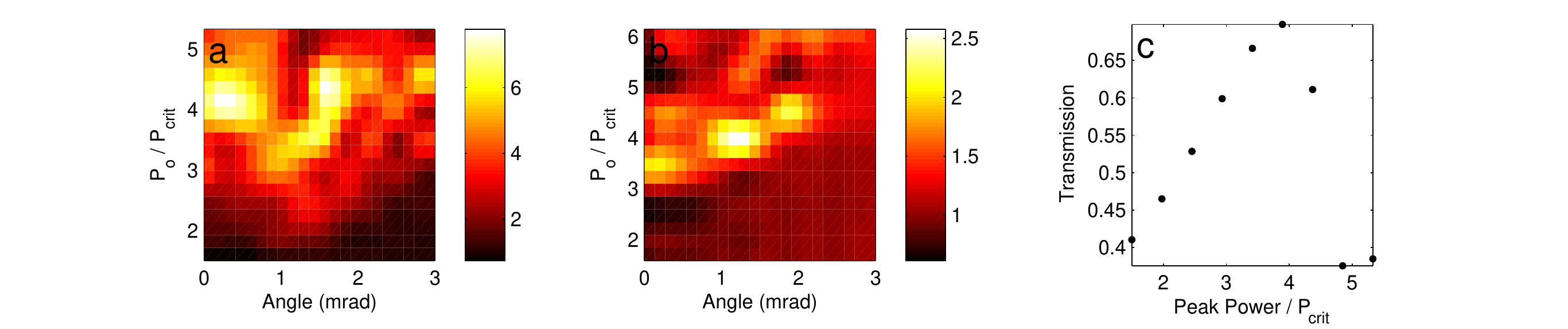}}
\caption{\textbf{Simulation of parameter dependence of gate operation for transform-limited, flat-phase Gaussian pulses with initial 200 $\mu$m waist and 320 fs duration (FWHM).} Plots show approximate interaction strength: relative amount of energy transferred off the central beam axis, varying pump power and relative angle equal energy signal, pump transmission through a 250 $\mu$m square aperture relative to transmission without signal present. Plotted data was interpolated by a factor of two along each axis. \textbf{a,} for gain = 1, \textbf{b} for gain =5 (power axis is pump peak power). \textbf{c,} Transmission of pump beam, without signal, through square 250 $\mu$m aperture immediately after propogation through 30 cm SF11. The eventual decrease of transmission (decreased localization of beam energy) with power indicates the point where collapse occurs within the material.
}
\label{symbr}
\end{figure}

We then introduced the signal beam at a small fraction of the pump power and varied the relative angle between the beams, for each angle scanning the time delay through zero. The signal beam perturbs the pump beam by causing it to collapse quicker, but also by changing the propagation of the pump through and beyond the collapse. Accordingly, we observed strong interactions not only for angles near zero, but also at angles for which strong soliton-like effects were observed with equal beam powers (see Supplementary Figure 2 a-f, Figure 2 in the letter). It is also possible to observe soliton-like interactions with a weak signal, as in Supplementary Figure 2 g-h. These occur only for relatively large angles and do not provide significant net deflection. These interactions could be used to control the pump optically without causing collapse. 

Supplementary Figure 5a shows the ratio of energy transmitted through a 250 $\mu$m square aperture without the signal present to that without the signal present for gain =1 (\textit{i.e.}, $T_{no signal}/T_{with signal}$). Supplementary Figure 5b shows the same for gain =5. Note the slightly different y-axis in the plots. These plots show roughly the extent to which the signal's presence causes the pump to transform, measured by the change in spatial energy distribution of the pump beam at the end of the SF11 rod. For equal energies, very strong interactions occur for particular non-zero angles, which increase with the power in the beams. When gain is equal to 5, relatively strong effects are still possible at these angles. In order to optimize the gate to be cascadable however, one must restrict the pump power below that for which collapse occurs in the gate without the signal present. Supplementary Figure 5c shows the transmission of the pump through the aperture without any perturbing signal beam. Beyond about 4$P_{crit}$ (1.36 MW), the pump's transmission decreases with increasing power. This 'point of no return' indicates that collapse has occurred in the material. During collapse, the pump will typically be affected by higher-order nonlinearities, as well as reshaping involving strong coupling between different effects. These effects make linear regeneration impossible. Hence, we operate well below this value and choose a relative angle within a range of strong interaction, about 1.2 mrad. Other operating points are possible. At a minimum, the signal must cause collapse to occur within the gate and additional logic contrast can be obtained using non-zero angles for which the pump's spatiotemporal instability is most strongly excited during and beyond the collapse. However, provided collapse is induced by the signal and does not occur when the signal is absent, logic contrast will be enhanced with propagation through subsequent gates, linear regenerators and, if necessary, additional linear or nonlinear propagation. 

\end{document}